\documentclass[aps,prl,twocolumn,showpacs,floatfix]{revtex4-1}
\usepackage{amsmath,bm,epsfig}
\usepackage{epsfig,color}
\usepackage{bm} 
\usepackage{ulem}

\newcommand{\bx}[1]{{\bm{x}}}

\renewcommand{\it}[1]{\textit{#1}}

\newcommand{\REM}[1]{}

\begin{document}
\title{Cumulative compressibility effects on population dynamics in
  turbulent flows}

\author{Prasad Perlekar$^{1,3}$, Roberto Benzi$^{2}$, David
  R. Nelson$^{3}$ and Federico Toschi$^{1,4}$}

\affiliation{$^{1}$Department of Physics, and Department of
  Mathematics and Computer Science, and J.M. Burgerscentrum, Eindhoven
  University of Technology, 5600 MB Eindhoven, The Netherlands; \\and
  International Collaboration for Turbulence Research. \\
  $^{2}$ Dipartimento di Fisica and INFN, Universit\`a ‘‘Tor Vergata’’, Via della Ricerca Scientifica 1, I-00133 Roma, Italy. \\
  $^{3}$ Lyman Laboratory of Physics, Harvard University, Cambridge,
  MA 02138, USA.\\$^{4}$ IAC, CNR, Via dei Taurini 19, 00185, Roma,
  Italy.}

\begin{abstract}
  Bacteria and plankton populations living in oceans and lakes
  reproduce and die under the influence of turbulent
  currents. Turbulent transport interacts in a complex way with the
  dynamics of populations because the typical reproduction time of
  microorganism is within the inertial range of turbulent time
  scales. In the present manuscript we quantitatively investigate the
  effect of flow compressibility on the dynamics of populations. While
  a small compressibility can be induced by several physical
  mechanisms, like density mismatch or the finite size of
  microorganisms with respect to the fluid turbulence, its effect on
  the the carrying capacity of the ecosystem can be dramatic. We
  report, for the first time, how a small compressibility can produce
  a sizeable reduction in the carrying capacity, due to an integrated
  effect made possible by the long replication times of the organisms
  with respect to turbulent time scales. A full statistical
  quantification of the fluctuations of population concentration field
  leads to data collapse over a broad range in parameter space.
\end{abstract}

\pacs{47.27.-i, 47.27.E-, 87.23.Cc \hspace{0.5cm} Keywords: Fisher
  equation, population dynamics, turbulence}

\maketitle
Bacterial colonies living on a nutrient-rich hard agar dish are
commonly described in terms of the
Fisher-Kolmogorov-Petrovsky-Piscounov (FKPP) equation of population
dynamics \cite{fisher}. On a rigid substrate the FKPP equation
includes only diffusion and growth terms. In a fluid environment the
flow can play an important role by means of the advective transport of
microorganisms, both at low \cite{Wakika}, as well as at high Reynolds
numbers. In particular, in oceans and lakes microorganisms have found
ways to thrive and prosper in high Reynolds number fluid environments.

As discussed in \cite{1d,per10}, the advection by means of a
compressible turbulent flow can lead to highly non-trivial dynamics.
Two striking effects emerge: for small enough growth rate, $\mu$, the
population concentration field, $c({\bm x},t)$, is strongly localized
near the transient but long-living sinks of the turbulent flows; in
the same limit, the average space-time concentration of the population
(carrying capacity) becomes much smaller than its maximum value in
absence of flow.  Both effects have important biological implications.
Recently it was shown that this phenomenology holds true in the case
of more realistic two dimensional compressible turbulent velocity
fields \cite{per10}.

In this manuscript we present new results that support the key role
played by even a slight level of compressibility.  We have in mind a
simplified model for photosynthetic microorganisms that actively
control their buoyancy to maintain themselves at a fixed depth below
the surface of a turbulent fluid.  Small density mismatches or
inertial effects due to the finite size of the microorganisms can also
lead to an effectively compressible flow \cite{tos09}.

The main result of our investigation is that the overall carrying
capacity can be described as a universal function of the
non-dimensional growth rate and compressibility. Our results imply
that even a small compressibility leads to a substantial reduction in
the overall carrying capacity.

\label{model}
Bacterial populations in nutrient rich environments can be 
modeled in terms of the following Fisher-Kolmogorov-Petrovsky-Piscunov (FKPP) equation:
\begin{equation} 
  {\frac{\partial c}{\partial t}} + {\bm \nabla}
  \cdot ({\bm u}c) = D \nabla^2 c + \mu c(1 - c).
\label{eq:fish}
\end{equation}
The above equation describes the evolution of the microorganisms
concentration field, $c({\bm x},t)$, under the effect of advective
transport by means of a (turbulent) velocity field, ${\bm u}({\bm
  x},t)$, and in presence of spatial diffusivity with a diffusion
constant, $D$, and a replication rate, $\mu$. The parameters $D$ and
$\mu$ encode important biological information which may depend on the
amount of nutrients, temperature, organismic motility and many other
parameters of the ecosystem.  In Eq. \eqref{eq:fish} the averaged
population density in absence of flow which can be sustained by the
ecosystem has been rescaled to $1$; hence our concentration field $c$
is dimensionless. As an example of ``life at high Reynolds number''
one could consider the FKPP equation [Eq.~\eqref{eq:fish}] to
represent the density of the marine cyanobacteria Synechococcus
\cite{moo95} under conditions of abundant nutrients (so that $\mu \sim
constant$).

\begin{table}[!t]
  \centering
  \begin{tabular}{@{} ccccccc @{}}
    \hline
    $\theta$ & $\kappa$ & $E$ & $\varepsilon$ & $\tau_{\eta}$ & $Z_{\mu \to 0}$  \\ 
    \hline
    $0$ & $0.0$ & $3.17$ & $0.61$&$5.80\cdot 10^{-2}$  & $1.0$  \\ 
    $15$ & $1.45\cdot 10^{-2}$ & $3.14$ &$0.55$& $6.10\cdot 10^{-2}$ &  $6.8\cdot 10^{-1}$   \\ 
    $30$ & $6.50\cdot 10^{-2}$ & $2.76$ &$0.47$& $6.60\cdot 10^{-2}$   &  $3.2 \cdot 10^{-1}$\\
    $45$ & $1.62\cdot 10^{-1}$ & $2.24$ &$0.36$& $7.60\cdot 10^{-2}$ & $1.6\cdot 10^{-1}$ \\
    $60$ & $3.90\cdot 10^{-1}$ & $1.68$ &$0.25$& $8.90\cdot 10^{-2}$  & $4.1 \cdot 10^{-2}$ \\ 
    $75$ & $7.50\cdot 10^{-1}$ & $1.38$ &$0.15$& $1.16\cdot 10^{-1}$  & $8.4\cdot 10^{-3}$ \\
    $90$ & $1.0$ & $1.25$ &$0.12$& $1.31\cdot 10^{-1}$ & $1.6\cdot 10^{-3}$ \\ 
    \hline
    \end{tabular}
    \caption{
      \label{tab:t1} For different values of $\theta$, the compressibility $\kappa$, the energy $E \equiv \frac{1}{L^2}\int u^2      {\rm d}x {\rm d}y$, the energy dissipation rate
      $\varepsilon\equiv \frac{\nu}{L^2}\int (\nabla {\bm u})^2 {\rm d}x {\rm d}y$, the Kolmogorov  time scale $\tau_{\eta}$ used in our simulations. 
      Different values of the compressibility are obtained by projecting out the compressible and incompressible component of our $2d$ compressible field for 
      $\kappa=0.16$ as discussed in the text. The energy and dissipation are rounded off to two decimal places.}
\end{table}
 
\label{results}
In this study we build a simple model of microorganisms living in a
localized layer at some depth e.g. under the ocean surface. For this
reason we will consider a two-dimensional planar surface, taken out of
a fully three dimensional turbulent flow. A projected two dimensional
slice from a fully 3d velocity field is compressible. The
dimensionless numbers characterizing the evolution of the scalar
field, $c({\bm x},t)$, are the Schmidt number $Sc=\nu/D$ and the
dimensionless growth rate $\mu\tau_{\eta}$.  Here
$\tau_{\eta}\equiv(\nu/\varepsilon)^{1/2}$ is the Kolmogorov
dissipative time scale and $\varepsilon$ is the energy dissipation
rate of the fluid.  A particularly interesting regime arises when the
growth time $\tau_g\equiv \mu^{-1}$ lies in the middle of the inertial
range. Although many factors can affect the estimates, this is the
case for oceanic cyanobacteria and phytoplankton living in an
turbulent ocean. Indeed, oceanic turbulence has eddies with turnover
times ranging from minutes to months while typical microorganism
doubling times range from half an hour to half a day
\cite{mck09,mar03}.

\begin{figure}[!t]
\vspace{-0.2cm}
\begin{center}
  \includegraphics[width=\hsize]{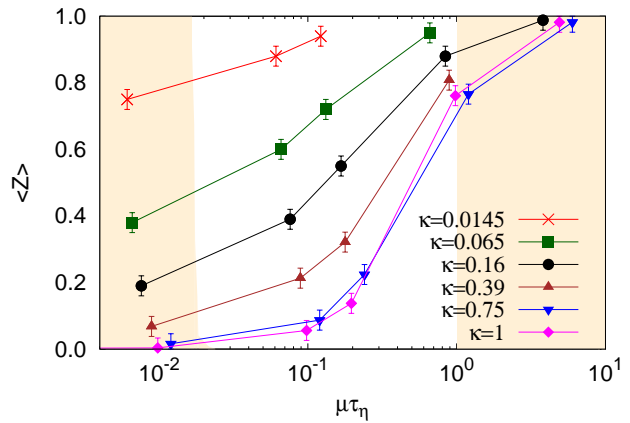}
\end{center}
\vspace{-1cm}
\caption{\label{fig:fig1}Time-averaged carrying capacity $Z$ versus
  the non-dimensional growth rate $\mu \tau_{\eta}$ for different
  values of compressibility $\kappa$. Region with
  $\mu\tau_{\eta}<\tau_{\eta}/\tau_{L}$ and $\mu\tau_{\eta}>1$ has
  been shaded for visual aid. Here, $\tau_L \equiv L/\sqrt{E}$ is the
  box size time scale with $L=2\pi$~\cite{per10}.}
\label{Z}
\end{figure}

\label{numerical}
Details of the numerical simulations can be found in \cite{per10}.
The velocity time series of the two dimensional slice were filtered in
order to produce time histories with different degrees of
compressibility. By projecting the velocity field into an
incompressible, ${\bm u}^{i}$, and compressible part, ${\bm u}^{c}$,
we could define a family of velocity fields ${\bm u}\equiv
\sqrt{2}[{\bm u}^{c} \sin(\theta) + {\bm u}^{i} \cos(\theta)]$ with $0
\leq \theta \leq \pi/2$. The dimensionless compressibility of each new
flow is $\kappa\equiv (\nabla\cdot {\bm u})^2/(\nabla {\bm u})^2$.
For $\theta=0$ the velocity field is purely incompressible,
$\kappa=0$, whereas for $\theta=\pi/2$ it is a potential flow,
$\kappa=1$. The case $\theta=\pi/4$ corresponds to the flow studied in
\cite{per10}. In Table~\ref{tab:t1} we report the values of the
compressibility used.
\label{sec:carc}
The carrying capacity of the ecosystem is defined as:
\begin{equation}
  Z=\langle \frac{1}{L^2}\int c({\bf x},t) \rm{d^2x} \rangle,
  \label{eq:carc}
\end{equation}
where the integral runs over a square domain of size $L^2$ with
periodic boundary conditions and $\langle \rangle$ indicates temporal
averaging. In Fig.~\ref{fig:fig1} we report $Z$ as a function of the
dimensionless growth rate, $\mu \tau_{\eta}$, for different values of
the flow compressibility (we recall that in absence of flow
$Z=1$). The shaded region on the right represents $\mu\tau_{\eta} >
1$, while the left shaded region reflects the condition
$\mu\tau_{\eta} < \tau_{\eta}/\tau_L$, where $\tau_L=L/\sqrt{E}$ is
the large eddy turnover time.  For large growth rates the curves
saturate towards unity, although for smaller compressibility the drop
in carrying capacity shifts to smaller and smaller growth rates,
$\mu\tau_{\eta}$.  Note that even at very small values of the
compressibility (e.g. $\kappa=0.0145$) there is a significant
reduction of the carrying capacity, $\sim 20\%$, for small growth
rates.  This regime is particularly relevant to marine biology where
compressibility is small but the orgamisms have a very slow
reproduction rate when compared with turbulent time scales. As a
result, despite small compressibility, one should still expect
important effects on the global carrying capacity.
For large values of the growth rate, i.e. $\tau_g\ll \tau_\eta$, the
time-scales of turbulence are too slow to have any effect on the
evolution of the concentration, and $Z\to 1$.

In the limit $\tau_g \to 0$, the concentration field tends to become
uniform with the leading correction coming from the local
compressibility. After a series expansion one obtaines $\langle
Z\rangle_{\mu} \approx 1 - (\tau_g^2/L^2) \langle \int ({\bm \nabla}
\cdot {\bf u})^2 {\rm {d \bf x}}\rangle + {\cal O}(\tau_g^3)$.  The
limiting values for {\it {small}} growth rates are complicated to
access numerically, especially for $\kappa<0.16$, as it takes a longer
and longer time for the carrying capacity to reach a steady state at
decreasing $\mu=\tau_g^{-1}$. To obtain reliable estimates in this
limit we can proceed as suggested in ~\cite{per10}
\begin{figure}[!t]
  \begin{center}
  \includegraphics[width=\hsize]{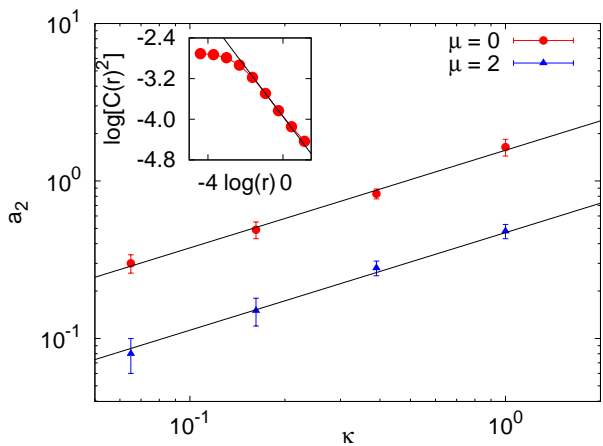}
  \end{center}
  \vspace{-0.7cm}
  \caption{Log-log plot of the anomalous scaling exponent $a_2$ as a
    function for the compressibility $\kappa$ in the case $\mu=0$ and
    $\mu=2$. Note that the two scaling laws have the same slope with
    different offset. Inset: Scaling behavior of $\langle C^2_\mu(r)
    \rangle$ for $\mu=0$ and $\kappa=0.16$.}
\label{fig_a2}
\end{figure}
where it was shown using a multifractal analysis that as $\mu\to 0$
the statistics of $c$ closely resembles the statistics of the
corresponding passive scalar probability density $P({\bm x},t)$, which
satisfies Eq.~\eqref{eq:fish} with $\mu=0$.  The limiting values of
$Z_{\mu \to 0}$ for different values of compressibility is given in
Table~\ref{tab:t1}.

\label{sec:univ}

The spatial behavior of $c({\bm x},t)$ is characterized by strong
fluctuations with non trivial correlations in both space and time.  In
\cite{per10} it was shown that the multifractal analysis describes the
statistical properties of the concentration field. Here we extend the
analysis to different values of the compressibility, $\kappa$, and of
the growth rate, $\mu$.  We focus on the scaling of:
\begin{equation}
  C^{q}_{\mu} (r)  = \frac{1}{r^2} \left<\int_{B(r)} c(x,t)^q d^2x \right>
\label{cq}
\end{equation}
where $\langle \dots \rangle$ now stands for a time and ensemble
average and $B(r)$ is a disk of radius $r$. Within the inertial range,
we find that $C_{\mu}^q (r) \sim r^{-a_q}$ with $a_q$ a non linear
function of $q$ constrained to $a_0=a_1=0$ \cite{roberto, frisch} (in
the inset of Figure \ref{fig_a2} we show the scaling of $C_{\mu}^2(r)$
versus $r$ for a particular value of $\mu=0$ and $\kappa=0.16$; plots
for other values look similar).  It is important to remember that each
$\kappa > 0$ can be associated with a compressibility length (and
time) scale, $l_{\kappa} = u_{rms}/\langle {(\nabla \cdot u)}^{2}
\rangle^{1/2}$ where $u_{rms}=\sqrt{E}$ is the root-mean-square
velocity. For very small compressibilities, this scale will be
comparable or larger than the integral scale of turbulence and the
system can effectively be considered as incompressible. In our system
this happened roughly for values of $\kappa < 0.065$ and indeed for
these compressibility values no scaling can be detected, making it
impossible to compute $a_q$. Hereafter we limit our analysis to the
region $ \kappa \ge 0.065$.

We expect the scaling exponent $a_2$ to be a function of $\mu$,
$\tau_{\eta}$ and $\kappa$.  Using dimensional analysis, we expect
that $a_2$ depends on $\mu$ and $\tau_{\eta}$ through the
dimensionless combination $\mu \tau_{\eta}$ and that the time scale
associated to the velocity gradient in the turbulent flow with
compressibility $\kappa$, should be given by $1/\langle (\nabla \cdot
\bf u)^2 \rangle^{1/2}= \tau_{\eta}/ \sqrt{\kappa}$. Hence, we
conjecture that $a_2$ should depend only on the dimensionless
combination $\sqrt{\kappa}/\mu \tau_{\eta}$. Since for $\mu=0$ we find
that $a_2$ has a finite value for any value of compressibility
$\kappa$, we further conjecture that $a_2\propto
\sqrt{\kappa}/{(\mu\tau_{\eta}+\beta)}$ where $\beta$ is a constant.
We expect that $\beta \approx \tau_{\eta}/\tau_L \approx
1/{Re_{\lambda}}$, where $\tau_L$ is the large eddy turnover time
discussed above.

\begin{figure}[!t]
  \begin{center}
  \includegraphics[width=\hsize]{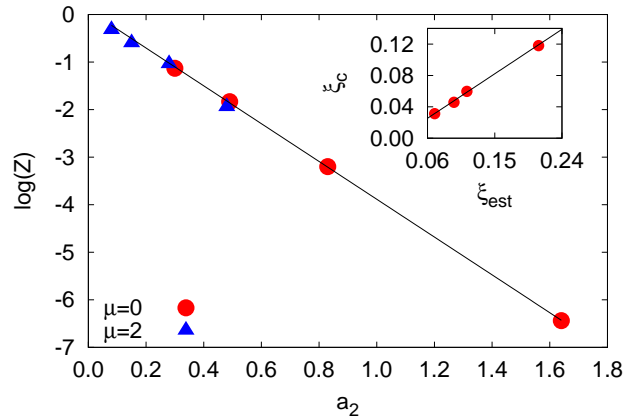}
  \end{center}
  \vspace{-0.8cm}
  \caption{Logarithmic plot of the carrying capacity $Z$ versus the
    anomalous exponent $a_2(\kappa)$ for $\mu=0$ and $\mu=2$. The
    black line represent a best fit with slope $-3.9\pm0.1$. We fit
    the data separately through $\mu=0$ and $\mu=2$ datasets. The mean
    of the fits is the slope and the error is the maximum deviation of
    the fit from the two slopes.  Inset shows the linear relationship
    between the localization length and the estimated cut-off length
    for $Sc=0.1,0.5,1$, and $10$. The black line represents the best
    fit and has a slope $\approx 0.6$. }
\label{fig_mu01}
\end{figure}

In Fig.~\ref{fig_a2}, we show the scaling exponent $a_2$ for $\mu=0$
and $\mu=2$ and for different compressibilities $\kappa$.  The most
striking feature of Fig.~\ref{fig_a2} is the well defined scaling law
between $a_2$ and $\kappa$ and that the scaling exponent does not
change by changing $\mu \tau_{\eta}$. Therefore we can write:
\begin{equation}
  a_2(\kappa, \mu \tau_{\eta}) = a(\mu \tau_{\eta}) \kappa^{\gamma}
\label{a2}
\end{equation}
where $a(\mu \tau_{\eta})=\alpha/(\mu\tau_{\eta}+\beta)$.  Using the
data in Fig.~\ref{fig_a2} we obtain $\gamma=0.62$, reasonably close to
the prediction $\gamma = 0.5$ given by dimensional analysis. Note that
the function $a(\mu \tau_{\eta})$ in Eq.~\eqref{a2} is a decreasing
function of $\mu \tau_{\eta}$, and that $\lim_{\mu \tau_\eta \to 0}
a_2(\mu\tau_\eta) = a_2(0) > 0$.

In Ref.~\cite{per10} it was shown that for $\mu=0$, the carrying
capacity is related to $a_2$ by $Z\approx (\xi_0/L)^{a_2}$ where
$\xi_0^2\equiv \langle P^2 \rangle/\langle (\nabla P)^2 \rangle$ and
$P({\bm x},t)$ is the solution of Eq.~\eqref{eq:fish} with $\mu=0$.
Here we generalize the results of Ref.\cite{per10} for $\mu\neq 0$ and
compressibilities $\kappa$ (see Table.~\ref{tab:t1}). In
Fig.~\ref{fig_mu01} we plot $\log(Z)$ versus
$a_2(\mu\tau_{\eta},\kappa)$ for $\mu=0$ and $\mu=2$, which supports
the scaling ansatz
\begin{equation}
  Z (\mu, \tau_{\eta}, \kappa ) =  \left(\frac{\xi_{est}}{L}\right)^{ a_2(\mu \tau_{\eta},\kappa)} 
\label{univfunc}
\end{equation}
which defines the length scale $\xi_{est}$. In order to gain a deeper
physical insight into the meaning of the cutoff scale $\xi_{est}$ we
define a scale based on the gradient of the concentration field which
is a generalization of our previous definition of $\xi_0$:
\begin{equation}
  \xi_c^2 = \frac{\langle c^2 \rangle }{ \langle (\nabla 
    c )^2 \rangle}.
\label{}
\end{equation}
In the inset of Fig.~\ref{fig_mu01} we show the behavior of the two
length scales $\xi_{est}$ and $\xi_c$ by varying the Schmidt number
$0.1\leq Sc \leq 10$.  Evidently the two definitions are consistent
within a numerical prefactor, suggesting that $\xi_{est}$ is
proportional to the cutoff scale of the gradients of the concentration
field.

We can now test for a universal scaling behavior of the carrying
capacity for varying compressibility $\kappa$ and non-dimensional
growth rate $\mu\tau_{\eta}$ using Eq.~\ref{univfunc}.

\begin{figure}[!t]
\begin{center}
  \includegraphics[width=\hsize]{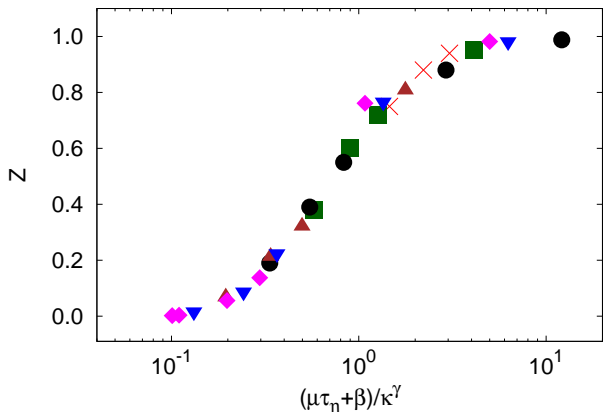}
\end{center}
\vspace{-1cm}
\caption{ Semi-logarithmic plot of $Z$ as a function of $(\mu
  \tau_{\eta}+\beta)/ \kappa^{\gamma} $ where $\gamma=0.62$ is given
  by the fit of Fig.~\ref{fig_a2} and $\beta=0.1$. All the data of
  Fig.~\ref{Z} has been collapsed on a single curve.}
\label{fig_Z_new}
\end{figure}

Upon defining the quantity $X \equiv \log(a(\mu
\tau_{\eta})\kappa^{\gamma}) $ the validity of Eq.~\eqref{univfunc}
can be checked by replotting the data of Fig.~\ref{Z} for $Z$ as a
function of $X$. Since $X = \log(\alpha)+ \log[\kappa^{\gamma}/(\mu
\tau_{\eta}+\beta)]$ and $\gamma=0.62$ from Fig.~\ref{fig_a2}, data
collapse of the results shown in Fig.~\ref{Z} should hold for some
particular value of $\beta$ whereas the exact value of $\alpha$
represents just a horizontal shift in the whole data set. In
Fig.~\ref{fig_Z_new} we show that for $\beta=0.1$ all the data
collapse on a well defined curve which represents our universal
function. This estimate $\beta=0.1$ obtained for the data collapse is
very close to the dimensional estimate $\beta\approx
\tau_{\eta}/\tau_{L} \approx 0.02-0.03$ (shaded region on the left in
Fig.~\ref{fig:fig1}).  This result is important since it shows the
deep and highly non-trivial connection between a bulk property of the
system, namely $Z$, and the intermittency parameters of the FKPP
equation in compressible flows. Moreover, Fig.~\ref{fig_Z_new}
represents a prediction of the average carrying capacity covering an
entire parameter space spanned by the three basic quantities of the
system, namely $\mu$, $\kappa$ and $\tau_{\eta}$. Our numerical
results can thus be extrapolated to new regimes, to investigate the
importance of weak compressibility due, for instance, to inertial
effects or buoyancy forces in populations subject to oceanic
turbulence.

In summary, we have used the FKPP equation to numerically study the
population dynamics in a compressible turbulent velocity field. As a
simplified model relevant for marine biology we have considered
microorganisms (bacteria or plankton) confined to a two-dimensional
plane experiencing the effect of a three dimensional velocity
field. We have investigated in detail the effect of compressibility in
a velocity field over a wide range of $\kappa$'s.  We found that even
a small compressibility can significantly reduce a global quantity
like the average carrying capacity, due to the slow reproduction rate
of the organisms. We expect that in oceans or lake this may be a
common situation. We further quantified in terms of spatial
intermittency exponents the statistical properties of the
concentration field.  Our study clearly suggests that it is
quantitatively wrong to neglect even small degrees of effective
compressibility. This compressibility, even in absence of organisms
with active buoyancy control, can be induced by density mismatches or
by the finite size of the organisms. Experimental tests of our
findings would be of primary importance.

{\bf Acknowledgment} We thank L. Biferale, H.J.H. Clercx, M.H. Jensen
and S. Pigolotti for useful discussions.  We acknowledge computational
support from CASPUR (Roma, Italy under HPC Grant 2009 N. 310), from
CINECA (Bologna, Italy) and SARA (Amsterdam, The Netherlands). Support
for D.R.N. was provided in part by the National Science Foundation
through Grant No. DMR-1005289 and by the Harvard Materials Research
Science and Engineering Center through NSF Grant DMR-0820484.  We
acknowledge the COST Action MP0806 for support. PP and FT acknowledge
the Kavli Institute of Theoretical Physics for hospitality. This
research was supported in part by the National Science Foundation
under Grant No. NSF PHY05-51164.  Data from this study are publicly
available in unprocessed raw format from the iCFDdatabase
(http://cfd.cineca.it).

\end{document}